\documentstyle[12pt,epsf]{article}
\def\Slash#1{{\rm\ooalign{\hfil$/$\hfil\crcr \hbox{$#1$}}}}

\newcommand\Kmbf[1]{\mbox{\boldmath{$#1$}}}
\newcommand\Kmbff[1]{\mbox{\footnotesize\boldmath{$#1$}}}

\begin{document}

\newcommand{\BQ}{\begin{equation}}
\newcommand{\EQ}{\end{equation}}
\newcommand{\BQA}{\begin{eqnarray}}
\newcommand{\EQA}{\end{eqnarray}}
\newcommand{\be}{\begin{eqnarray}}
\newcommand{\ee}{\end{eqnarray}}
\newcommand{\NN}{\nonumber \\}
\newcommand{\del}{\partial}
\newcommand{\tr}{{\rm tr}}
\newcommand{\Tr}{{\rm Tr}}
\newcommand{\Path}{{\rm P}\,}
\newcommand{\ket}[1]{\left.\left\vert #1 \right. \right\rangle}
\newcommand{\bra}[1]{\left\langle\left. #1 \right\vert\right.}
\newcommand{\ketrm}[1]{\vert {\rm #1} \rangle}  
\newcommand{\brarm}[1]{\langle {\rm #1} \vert}  
\newcommand{\V}{\widetilde V}
\newcommand{\x}{x_\perp}
\newcommand{\y}{y_\perp}
\newcommand{\z}{z_\perp}
\newcommand{\q}{q_\perp}
\newcommand{\p}{p_\perp}
\newcommand{\kk}{k_\perp}
\newcommand{\N}{{\cal N}_\tau}
\newcommand{\e}{\epsilon}
\newcommand{\al}{\alpha}
\newcommand{\la}{\lambda}
\newcommand{\K}{{\cal K}}
\newcommand{\bb}{b_\perp}
\newcommand{\r}{r_\perp}

\begin{flushright}
SACLAY-T03/115\\
hep-ph/0308096 
\end{flushright}

\begin{center}
\vspace{3cm} 
{\LARGE\bf
Vector Mesons on the Light Front}
\vspace{2mm}
\\

\vspace{1.5cm}

{\large
 K. Naito$^{\, a,}$\footnote{\tt knaito@nucl.sci.hokudai.ac.jp}, 
S. Maedan$^{\, b,}$\footnote{\tt maedan@tokyo-ct.ac.jp} and  
K. Itakura$^{\, c,d,}$\footnote{\tt itakura@quark.phy.bnl.gov, 
itakura@spht.saclay.cea.fr} 
}
\\
\vspace{1.3cm}
{\it $^a$ 
    Meme Media Laboratory, 
    Hokkaido University, Sapporo 060-8628, Japan}\\
{\it $^b$ Department of Physics, Tokyo National College of Technology, 
Tokyo 193-0997, Japan}\\
{\it $^c$ RIKEN~BNL~Research~Center, Brookhaven National Laboratory, Upton, 
NY 11973, USA}\\
{\it $^d$ Service de Physique Th\'eorique, CEA/Saclay, F91191 
Gif-sur-Yvette Cedex, France \footnote{present address}}
\vspace{13mm}
\end{center}

\vspace{.3cm}

\begin{abstract}
We apply the light-front quantization to the Nambu--Jona-Lasinio model 
with the vector interaction, and compute vector meson's mass and light-cone
wavefunction in the large $N$ limit. Following the same procedure
as in the previous analyses for scalar and pseudo-scalar mesons, 
we derive the bound-state equations of a $q\bar q$ system in the 
vector channel. We include the lowest order effects 
of the vector interaction. The resulting transverse and longitudinal 
components of the bound-state equation look different from each other.
But eventually after imposing an appropriate cutoff, one 
finds these two are identical, giving the same mass and 
the same (spin-independent) light-cone wavefunction. Mass of the vector 
meson decreases as one increases the strength of the vector interaction.
\end{abstract}

\newpage

The light-cone (LC) wavefunction of a hadron is one of the most useful
quantities for describing the hadron structure in terms of its underlying
degrees of freedom \cite{Brodsky_lecture}.  In general, it contains 
information about soft dynamics
among quarks, antiquarks and gluons, and one can compute various scattering 
processes involving hadrons in initial/final states, by 
combining it with the hard part of the diagrams. 
Diffractive vector meson production 
is one of the typical examples of such processes \cite{Diffractive}. 
To compute the amplitude of diffractive electro/photoproduction of 
a vector meson for a wide range of kinematics, one needs to know 
the LC wavefunction of a vector meson with non-perturbative 
information. In this Letter, we are going to discuss this LC 
wavefunction of the vector meson in a simple model. 
As another interesting example, E791 experiment 
at Fermilab \cite{E791} has recently attempted to determine the LC 
wavefunction (squared) of pions through the diffractive pion 
dissociation process (dijets production) according to 
Ref.~\cite{pion_dissociation}. Although it is argued 
that determination of the pion LC wavefunction 
from the experimental data is actually quite hard \cite{Braun_Chernyak}, 
it is still true that one cannot compute the amplitude of this 
process without knowing the pion LC wavefunction.  Similar experiments 
are possible in the dijets production from a virtual photon, where,
according to the vector meson dominance model,  
the vector meson contribution forms the hadronic part of the photon
wavefunction.

Perturbative calculations provide us with the so called "asymptotic" form
of the LC wavefunctions. For example, the asymptotic form of the pion
LC wavefunction is known as well as the vector meson's one 
\cite{Brodsky-Lepage, Braun-Koike-Tanaka}. However, non-perturbative 
study is quite few. Lattice simulation can compute the 
first few moments of meson's distribution function, 
but at present they are not sufficient to determine the LC wavefunction 
itself. Therefore, it is very important to develop a 
non-perturbative technique which allows us to directly obtain the 
LC wavefunction. Clearly, the most straightforward and natural 
framework is the Hamiltonian formalism in the light-front (LF) 
quantization \cite{LF_review, Brodsky_lecture}. 
Before challenging the problem in the real QCD, 
one should be able to learn much from the analyses of simpler models 
such as the Nambu--Jona-Lasinio (NJL) model.
Indeed, this model was recently studied by two of the authors within the 
LF quantization \cite{Itakura-Maedan, Itakura-Maedan_review} 
and we follow the same procedure to get the LC wavefunctions
of vector mesons. As is well known, there is a paradoxical situation in the
LF quantization. It has been asked how one can describe spontaneous 
symmetry breaking  in a formalism having only a trivial Fock vacuum. 
This was answered in Ref.~\cite{Itakura-Maedan} within the 
NJL model with $N$ component fermions
$ {\cal L}_{\rm NJL} = \bar{\Psi} (i\Slash{\partial}-m^{}_0) \Psi
 + \frac{1}{2} G_1 [ (\bar{\Psi}\Psi)^2 + (\bar{\Psi}i\gamma_5
 \Psi)^2 ]$.
Based on the analogy with the { description} of spontaneous 
symmetry breaking in a scalar theory on the LF, they found that, 
still with the trivial Fock vacuum, chiral symmetry breaking is 
described in such a way 
that one selects an appropriate Hamiltonian depending on the phases of 
the symmetry. In the NJL model, different Hamiltonians 
are originated from different solutions to the constraint equation, 
which exists only in the LF formalism. The "bad" component of 
the spinor, $\psi_-$ 
(where $\psi_\pm=\frac12 \gamma^\mp \gamma^\pm \Psi$ and 
$\psi_+$ is called "good"\footnote{Our notation is the 
following:
$x^\pm=(x^0\pm x^3)/\sqrt2,\ v^\mu= (v^+,v^-, v_\perp^i), 
\Kmbf{v}=(v^+, v^1_\perp, v^2_\perp)$ and 
$\partial_\pm=\partial / \partial x^\pm$.
We use $\mu,\nu$ for Lorentz indices of four vectors, $i,j$ for transverse 
coordinates $1,2,$ and  $\alpha,\beta$ for spinor indices.}) 
is not a dynamical variable and is subject to a constraint equation, 
as we will see below. This "fermionic constraint" is a nonlinear 
equation in the NJL model and leads to the "gap equation" for the 
chiral condensate if one adopts an appropriate cutoff.
Namely, using the parity invariant cutoff $|p^\pm | <\Lambda$, 
one gets
\begin{equation}
\frac{M-m^{}_0}{M}=\frac{G_1 N\Lambda^2}{4\pi^2}\left\{2-\frac{M^2}{\Lambda^2}
\left(1+ \ln \frac{2\Lambda^2}{M^2}\right)\right\} \, ,
\label{gap}
\end{equation}
where $M=m^{}_0 -G_1 \langle \bar\Psi\Psi\rangle $ 
is the dynamical mass of the fermion. 
When the coupling constant $\widetilde G_1=G_1N\Lambda^2/4\pi^2$ 
is larger than the critical value $\widetilde G_1^{\rm (critical)}=1/2$, 
the gap equation has a non-zero solution even in the chiral limit 
$m^{}_0\to 0$. 
This means that the fermionic constraint allows for "symmetric" and
"broken" solutions corresponding to those of the gap equation. 
If one selects the "broken" solution, and substituting it to the canonical 
Hamiltonian, one obtains the "broken" Hamiltonian. This 
governs the dynamics in the broken phase and is completely 
different from the Hamiltonian with the "symmetric" solution.
In Ref.~\cite{Itakura-Maedan}, the fermionic constraint was solved 
by using the $1/N$ expansion [Indeed, Eq.~(\ref{gap}) is the leading 
order result], and they obtained the Hamiltonian in both symmetric
and broken phases. They also solved the bound-state equations
for the scalar and pseudo-scalar mesons\footnote{Though there was only 
one flavor in Ref.~\cite{Itakura-Maedan}, these physically 
correspond to the pion and sigma mesons, and also the vector meson
to be discussed in the present Letter corresponds 
to the rho meson. Generalization to multi flavors should be straightforward.} 
and obtained their 
LC wavefunctions and masses, as well as the PCAC and GOR relations. \\

One can of course apply the same procedure for vector states, but 
we know that the NJL model ${\cal L}_{\rm NJL}$ does not allow for
a bound state in the vector channel \cite{NJL_review, Klimt90}. 
Vector states start to bind if one
adds the vector interaction so that the attractive force between a quark and
an antiquark in the vector channel becomes stronger. 
Therefore, in this Letter, we 
include the vector interaction minimally by adding the following interaction:
\begin{equation}
{\cal L}_V = - \frac{G_2}{2} \left[ (: \bar{\Psi}\gamma_\mu\Psi :)^2
 + (: \bar{\Psi}\gamma_\mu\gamma_5 \Psi :)^2 \right]. \label{vector_int}
\end{equation}
This interaction, however, makes the fermionic constraint 
tremendously complicated:
\begin{eqnarray}
 \lefteqn{
 i\partial_- \psi_- = \left( i\gamma^i_\perp \partial_{\perp i} + m^{}_0\right)
 \frac{1}{2} \gamma^+ \psi_+ } \nonumber \\
 &&\hspace*{-4mm} {} - \frac{G_1}{2} \left[ 
  \gamma^+ \psi_+ (\bar{\psi}_+ \psi_- + \bar{\psi}_- \psi_+ ) 
 - i\gamma_5 \gamma^+ \psi_+
 (\bar{\psi}_+ i\gamma_5 \psi_- + \bar{\psi}_- i\gamma_5 \psi_+)
 \right]  
 \nonumber \\
 &&\hspace*{-4mm} {}+\hspace{0.3mm} G_2 \, \left[ \psi_- (:\bar{\psi}_+ \gamma^+ \psi_+:)
 + \gamma_5 \psi_- (:\bar{\psi}_+ \gamma^+ \gamma_5 \psi_+:) \right] 
 \nonumber \\
 &&\hspace*{-4mm} {} + \frac{G_2}{2} \left[
  \gamma^i_\perp \gamma^+ \psi_+
 :\left\{ \bar{\psi}_+ \gamma^i_\perp \psi_- +
 \bar{\psi}_- \gamma^i_\perp \psi_+ \right\}: 
 -  \gamma_\perp^i \gamma_5 \gamma^+ \psi_+
 :\left\{ \bar{\psi}_+ \gamma^i_\perp \gamma_5 \psi_- +
 \bar{\psi}_- \gamma^i_\perp \gamma_5 \psi_+ \right\}: 
 \right].  \label{LC:2}    
\end{eqnarray} 
Here we have followed the same operator ordering as in the previous 
analysis without the vector interaction. After rewriting this equation
into a bilocal form, one can solve it in the quantum level 
by using the $1/N$ expansion, which is systematically generated 
by the Holstein-Primakoff technique \cite{Itakura_BEM}.
It turns out that the leading order equation gives the same gap equation 
as Eq.~(\ref{gap}). This is natural because we have taken the normal 
order\footnote{Normal order is defined with respect to the Fourier modes
of the fermionic field.} in the interaction (\ref{vector_int}).

In the leading order of the $1/N$ expansion, mesonic states are 
written as constituent states with a (dynamical) quark and a (dynamical) 
antiquark, as was discussed in Ref.~\cite{Itakura-Maedan}
for the scalar and pseudo-scalar states. 
Thus, a generic vector state with total momentum $P^\mu$ and helicity 
$\lambda$ { can} be represented as follows:
\begin{eqnarray}
 | {\rm vector} ; \lambda, P \rangle\!\! &=&\!\! 
P^+ \int_{0}^{P^+} dk^+
 \int d^2 k_\perp \ \phi_\lambda (x,k_\perp)  
  \label{EQ:LCDEF} \\
&&\hspace*{-3mm}  \times\ \epsilon_\mu(P,\lambda) 
 \left\{ \Gamma^{\mu} ( -\Kmbf{k},-\Kmbf{P}+\Kmbf{k} )
 + \bar\Gamma^{\mu} (-\Kmbf{P}+\Kmbf{k},- \Kmbf{k}) \right\}_{\alpha\beta}
 B^\dagger_{\alpha\beta}(\Kmbf{k},\Kmbf{P}-\Kmbf{k})|0\rangle, \nonumber
\end{eqnarray}
where $k^+$ and $k_\perp$ are the  longitudinal and transverse momentum 
of the quark ($x=k^+/P^+$), 
$\phi_\lambda (x,k_\perp)$ is the spin independent part
\footnote{By definition, $\phi_\lambda(x,k_\perp)$ should be independent
of $\lambda$, but we retain $\lambda$ because, as we will see below, the 
bound-state equations look different for different $\lambda$.} of the 
LC wavefunction which should be determined by the dynamics, 
$\epsilon^\mu(P,\lambda)$ is a polarization vector,  
and $B^\dagger_{\alpha\beta}(\Kmbf{p},\Kmbf{q})$ is a bosonic operator
which was introduced to solve the fermionic constraint \cite{Itakura-Maedan} 
and corresponds to quark and antiquark creation operators 
$\sim b^{\alpha\dagger}_{\Kmbff{p}} d^{\beta\dagger}_{\Kmbff{q}}$
in the leading order of the $1/N$ expansion.
Spin dependent part of the LC wavefunction, 
$\Gamma^\mu=(\Gamma^+,\Gamma^-,\Gamma^i)$, 
is determined by the interpolating field of the vector meson:
\begin{eqnarray}
&& \Gamma^+(\Kmbf{p},\Kmbf{q}) =
 \bar\Gamma^+(\Kmbf{p},\Kmbf{q}) = \frac{1}{2}\cdot {\bf 1}\, ,
 \label{LCEQ:H141127:2}\\
&& \Gamma^-(\Kmbf{p},\Kmbf{q})  =  \bar\Gamma^- (\Kmbf{q},\Kmbf{p}) =
 -\frac{1}{4\, p^+q^+} \left\{ 
 \gamma^i_\perp p^i_\perp \gamma^j_\perp q_\perp^j + M
 \gamma^i_\perp (p_\perp + q_\perp)^i + M^2 \right\}\, , 
 \label{LCEQ:H141127:3} \\
&& \Gamma^{i}(\Kmbf{p},\Kmbf{q}) =  -\frac{1}{2 q^+} \gamma^i_\perp 
 \left(\gamma^j_\perp q_\perp^j + M \right) ,\qquad
 \bar\Gamma^{i}(\Kmbf{p},\Kmbf{q}) =  -\frac{1}{2 q^+}
 \left(\gamma^j_\perp q_\perp^j + M \right) \gamma^i_\perp\, . \label{LCEQ:H140907:4}
\end{eqnarray}
The polarization vector is written in  the rest frame of the meson
$P^\mu =
({m^{}_{\rm V}}/{\sqrt{2}},{m^{}_{\rm V}}/{\sqrt{2}},
\vec 0_\perp)$ as (see also \cite{Ji})
\begin{equation}
 \epsilon^\mu(P,\lambda=\pm 1) = 
 \left( 0,0,\frac{\mp 1}{\sqrt{2}},\frac{-i}{\sqrt{2}} \right),\quad
 \epsilon^\mu(P,\lambda=0) = \left( \frac{1}{\sqrt{2}} , 
   -\frac{1}{\sqrt{2}},\, \vec 0_\perp \right)  .
 \label{EQ:POL}
\end{equation} 
As we explained before, after solving the fermionic constraint 
(\ref{LC:2}) with the nontrivial
solution $M\neq 0$ to the gap equation (\ref{gap}) and substituting its
solution into the canonical Hamiltonian, one gets the "broken" 
Hamiltonian $H_{\rm LF}$.
 Its explicit form and the detailed derivation of the 
Hamiltonian are very complicated and will be reported elsewhere 
\cite{full_paper}. Here only the final results are shown. 
The Hamiltonian is derived order by order of the $1/N$ expansion, 
$$H_{\rm LF}=N\sum_{n=0}^\infty \left(\frac{1}{\sqrt N}\right)^n h^{(n)},$$
and $h^{(2)}$ turns out to be the lowest non-trivial Hamiltonian since 
$h^{(0)}$ is just a constant and $h^{(1)}=0$.
Therefore, keeping this non-trivial order,
one can write the eigenvalue equation for a vector state as
\begin{equation}
 h^{(2)} |{\rm vector};P \rangle = \frac{ m_{\rm V}^2 + P_\perp^2 }{2P^+ }
 | {\rm vector}; P \rangle. 
 \label{EQ:EV}
\end{equation}
Solving this equation yields both the spin independent part of 
the LC wavefunction $\phi_\lambda(x,k_\perp)$ and mass of the 
vector meson $m^{}_{\rm V}$ simultaneously. 

Before going into details, let us briefly discuss the 
$q\bar q$ states to clarify the procedure we perform.
In general, the LF energy of the two body state (\ref{EQ:LCDEF}) 
may be schematically written as 
\BQ
P^-_{q\bar q} = \frac{\kk^2+M^2}{2k^+} + 
\frac{(P_\perp-\kk)^2 + M^2}{2(P^+-k^+)} + V(k,P),
\EQ
where the first two terms are the "kinetic" energies of the quark and 
the antiquark, and $V$ is the potential which allows for a bound state.
This form of the energy leads to the following bound-state equation:
\BQ
\left\{m_{\rm V}^2 - \frac{\kk^2+M^2}{x(1-x)}\right\}\phi(x,\kk)=
\int_0^1dy \int d^2\p\, V(x,\kk;y,\p)\, \phi(y,\p)\, ,
\EQ
where we have chosen the vector meson's rest frame, 
$\Kmbf{P}= ( P^+, P_\perp^i )  = ( m_{\rm V} / \sqrt{2},\, 0_\perp ) $
 for simplicity, and redefined 
$V$ with some factors included. In the following, since we are 
interested in seeing how the vector interaction (\ref{vector_int}) 
affects the vector channel,  we will derive the potential $V$ 
up to the leading order of the vector interaction $G_2$. 
We will see that in this leading order 
the potential term $V(x,\kk;y,\p)$ is separable with respect to 
the internal $(y,\p)$ and external $(x,\kk)$ variables, 
and actually depends only on $y$ and $\p$. 
\\

Now let us explicitly show the bound-state equations of the transverse and 
longitudinal components derived from the leading nontrivial 
Hamiltonian $h^{(2)}$. 
First, for a transversely polarized vector meson, 
a lengthy calculation yields the following potential $V_{\rm T}$
($\epsilon(x) = x/|x|$)
\begin{eqnarray}
V_{\rm T}
 =  -\, \, \frac{ G_2 N  }{(2\pi)^3}  
\left[1  + \frac{G_2N}{(2\pi)^3} \int_{-\infty}^{\infty} dq^+
 \int d^2q_\perp \frac{ \epsilon(q^+) }{ P^+ - q^+ } \right]^{-1}
 \frac{ p_\perp^2 + M^2 - 2y(1-y) p_\perp^2 }{y^2(1-y)^2}\, .
\end{eqnarray} 
Notice that this is already 
independent of the external variables $x, \kk$. 
Taking the leading contribution of $G_2$, one arrives at an equation for 
the LC wavefunction $\phi^{}_{\rm T}(x,k_\perp)=
\phi_{\lambda=\pm 1}(x,k_\perp)$:
\begin{eqnarray}
\left\{  m_{\rm T}^2  - \frac{ k_\perp^2 + M^2}{x(1-x)} \right\} 
 \phi^{}_{\rm T}(x,k_\perp)  
 =  -\, \frac{ G_2 N  }{(2\pi)^3}  \int dy\, d^2 p_\perp
 \frac{ p_\perp^2 + M^2 - 2y(1-y) p_\perp^2 }{y^2(1-y)^2}\,
 \phi^{}_{\rm T}(y,p_\perp).
 \label{EQ:TRANS}
\end{eqnarray} 
Next, the longitudinal component is much more involved. 
A longer, but straightforward calculation leads to a more complicated 
potential $V_{\rm L}$:
\begin{eqnarray}
V_{\rm L} 
  &\! \! =&\!\! - \frac{ G_2 N }{(2\pi)^3  }
\left\{ 1 - 2 \frac{ G_2 N }{(2\pi)^3  } \int_0^1 dz \int d^2
q_\perp \right\}^{-1}      
 \left\{  m_{\rm L}^2  - \frac{ k_\perp^2 + M^2 }{x(1-x)} \right\} 
 \left\{  m_{\rm L}^2  + \frac{ k_\perp^2 + M^2 }{x(1-x)} \right\}^{-1} 
      \NN
 &  \times &\hspace*{-3mm}
\left[ \, 2 \,  
         +  \, \frac{ 4 \,( k_\perp^2+M^2 )
}{  m_{\rm L}^2  \, x(1-x) -(k_\perp^2+M^2 ) }
          \left\{ 1 - \frac{ G_2 N }{(2\pi)^3  } \int_0^1\! dz\! \int\! d^2
q_\perp \right\} \right]
   \left\{ m_{\rm L}^2  \, + \frac{ p_\perp^2 + M^2 }{y(1-y)} \, \right\}
  .     
         \label{EQ:STLONG}
\end{eqnarray}
Again, taking the leading term with respect to $G_2$ 
after careful modification of the bound-state equation, 
we eventually obtain the following simpler equation
for the longitudinal mode $\phi^{}_{\rm L}(x,k_\perp)=
\phi_{\lambda=0}(x,k_\perp)$:
\begin{eqnarray}
 \left\{  m_{\rm L}^2  - \frac{ k_\perp^2 + M^2 }{x(1-x)} \right\} 
 \phi^{}_{\rm L}(x,k_\perp) 
 =  - \, \frac{ G_2 N }{(2\pi)^3}
  \int dy\, d^2 p_\perp
 \frac{ 4(p_\perp^2 + M^2) }{y(1-y) }\, \phi^{}_{\rm L}(y,p_\perp)\, .
 \label{EQ:LONG}
\end{eqnarray} 
It is evident that the right-hand side is again 
independent of the variables $x,\kk$.
Since the step from Eq.~(\ref{EQ:STLONG}) to Eq.~(\ref{EQ:LONG}) 
is a bit nontrivial, let us show the easiest way to derive 
Eq.~(\ref{EQ:LONG}), which is however less systematic.
First of all, if one ignores the $G_2$ dependent term in the second line 
of Eq.~(\ref{EQ:STLONG}) that gives the higher order in $G_2$ and thus 
can be ignored anyway, one immediately finds 
$$
\left\{  m_{\rm L}^2  - \frac{ k_\perp^2 + M^2 }{x(1-x)} \right\} 
\left\{  m_{\rm L}^2  + \frac{ k_\perp^2 + M^2 }{x(1-x)} \right\}^{-1} 
\left[ \, 2 \,  +  \, \frac{ 4 \,( k_\perp^2+M^2 )}
                           {m_{\rm L}^2  \, x(1-x) -(k_\perp^2+M^2 ) }
\right]=2\, .
$$
Then one integrates the resulting bound-state equation over $x$ and $\kk$, 
obtaining the following:
$$
\int dx\, d^2\kk  \left\{  m_{\rm L}^2  - \frac{ k_\perp^2 + M^2 }{x(1-x)} \right\} 
 \phi^{}_{\rm L}(x,k_\perp) = -4 \frac{G_2N}{(2\pi)^3}
\left(\int dx\, d^2\kk\right) \int dy\, d^2\p \frac{\p^2+M^2}{y(1-y)}\, 
\phi^{}_{\rm L}(y,\p).
$$
One can modify the bound-state equation by using this integral in the 
right-hand side of it. Finally, taking the leading term with respect 
to $G_2$, one obtains Eq.~(\ref{EQ:LONG}).
It should be noted that the above equation is consistent with 
Eq.~(\ref{EQ:LONG})
since it is the integration of Eq.~(\ref{EQ:LONG}) over $x$ and $\kk$.

At first glance, the above two eigenvalue equations 
(\ref{EQ:TRANS}) and (\ref{EQ:LONG}) look different
and thus seem to give different masses for the transverse and longitudinal 
vector mesons.
This is of course physically unacceptable, and as we will verify soon, 
these equations are essentially the same and give the same mass 
$m^{}_{\rm T}=m^{}_{\rm L}$. This equivalence will be achieved 
after one specifies cutoff scheme.
It is not hard to identify the origin of this (fake) difference with 
the lack of Lorentz covariance in the LF formalism. 

Nevertheless, even at this stage, one can see that  solutions to
the above equations are an identical function of $x$ and $k_\perp$.
To this end, it should be noted again that these two equations have 
a very simple {structure}: Their right-hand sides  depend on neither
$x$ nor $k_\perp$.
This immediately implies that the solutions should be 
\begin{equation}
\phi^{}_{\rm T/L}(x,k_\perp)
 = \frac{C_{\rm T/L}}{m_{\rm T/L}^2 - \frac{k_\perp^2 + M^2}{x(1-x)}}\, .
 \label{EQ:SOL}
\end{equation}
Thus, if the masses of transverse and longitudinal vector mesons coincide 
with each other, then so do the LC wavefunctions ($C_{\rm T/L}$ are 
determined by the normalization). The LC wavefunction (\ref{EQ:SOL}) has 
a peak at $x=1/2$ as we will see explicitly below. 
Since our description of the vector states is with respect to 
the quark (antiquark) having a dynamical mass $M\neq 0$, 
this shape of the LC wavefunction implies the constituent picture.

Now let us verify that both the two equations
(\ref{EQ:TRANS}) and (\ref{EQ:LONG}) derive the same equation for a
vector meson mass $m^{}_{\rm V}$. Inserting the solution (\ref{EQ:SOL}) into 
these equations, one arrives at integral equations for $m^{}_{\rm V}$.
We evaluate the integrals 
by introducing the "extended parity invariant cutoff" 
\cite{Itakura-Maedan}
which is actually equivalent to the Lepage-Brodsky 
cutoff \cite{Brodsky-Lepage}: 
\begin{equation}
    \frac{  p_\perp^2 + M^2  }{ y(1-y)  } < 2 \, \Lambda^2\, .
    \label{LB}
\end{equation}
Indeed, this is a natural extension of the parity invariant cutoff 
in the {\it two body} sector, $K^+ K^- <\Lambda^2$ 
where $K^\pm$ are the sum of 
(on-shell) quark and antiquark longitudinal momenta and energies 
[$K^+=p^++(P^+-p^+)=P^+,\ K^-=(\p^2+M^2)/2p^+ + (\p^2+M^2)/2(P^+-p^+)$].
This cutoff { apparently} preserves transverse rotation and parity 
symmetry separately, but in fact it does work better. First, it also 
respects the usual three dimensional space rotation \cite{Bentz}. 
Thus one can relate the above cutoff $\Lambda$ to the 3-momentum 
cutoff $\sum_{i=1,2,3}(p^i)^2 < 
\Lambda^2_{\rm 3M}$ through $ 2(\Lambda^2_{\rm 3M}+M^2) = \Lambda^2$.
Next, the cutoff $K^+K^-<\Lambda^2$ is invariant under 
the boost transformation $K^\pm\to e^{\pm\beta}K^\pm$, 
which is necessary for the relativistic formulation.

 In Ref.~\cite{Itakura-Maedan}, the parity invariant cutoff was 
specified as $|K^\pm|<\Lambda$ which contains two independent conditions. 
In actual calculations, however, the authors of Ref.~\cite{Itakura-Maedan} 
utilized only the Lepage-Brodsky cutoff which is obtained by 
conbining the two conditions\footnote{Therefore, body of the calculations 
in Ref.~\cite{Itakura-Maedan} is correct, while the derivation of the 
Lepage-Brodsky cutoff was not appropriate.}.
Namely, the condition $|K^\pm|<\Lambda$ was introduced only to derive 
the Lepage-Brodsky cutoff. 
However, putting the cutoff on the longitudinal momentum 
$K^+<\Lambda$ for the total momentum is not preferable from the 
viewpoint of boost symmetry. Thus, in the present paper, we redefined 
the parity invariant cutoff in the two body sector by $K^+K^-<\Lambda^2$.
On the other hand, there is no problem in putting $|p^\pm|<\Lambda$ 
in the gap equation (1) because the momentum is not the external 
momentum but the internal one to be integrated out. Indeed, the gap 
equation comes from the zero longitudinal momentum of the fermionic
constraint written in the bilocal form.

Now the integral in the equations is replaced as follows\footnote{
Alternatively, one can regard that the LC wavefunction 
(\ref{EQ:SOL}) has support defined by the Lepage-Brodsky cutoff.}:
\begin{eqnarray}
\int dy  \int
 d^2 p_\perp \ \longrightarrow \ 
\int_{y_-}^{y_+} dy  \int_0^{2
\Lambda^2 y (1-y)-M^2 }  \pi  \, d (p_\perp^2) \, ,
  \end{eqnarray}
with 
\begin{equation}
  y_\pm = \frac{1 \pm \beta}{2},  \hskip1cm  \beta \equiv \sqrt{ 1- \frac{2
M^2}{\Lambda^2} }.\label{y_beta}
\end{equation}
Then one can explicitly prove that the two equations from 
Eqs.~(\ref{EQ:TRANS}) and (\ref{EQ:LONG}) do 
give the same equation that determines the mass 
$m^{}_{\rm V}=m^{}_{\rm T}=m^{}_{\rm L}$.  It is very important to 
recognize that we can derive this equation simply by inserting the LC 
wavefunction (\ref{EQ:SOL}) into the bound-state equations with the above 
cutoff. We have just evaluated the right-hand sides of the equations. 
Since the LC wavefunction (\ref{EQ:SOL}) is a direct consequence of the 
bound-state equations, to obtain the same equation for $m_{\rm V}$ means 
that the original equations are also equivalent to each other. 

The explicit form of the equation for the vector meson mass is given by 
\begin{equation}
   \frac{1}{ \widetilde G_2}
   =  \frac{2}{3} \left[ \, \beta  + (1- \beta^2) \left\{
        r \ln \left( \frac{1+\beta}{1-\beta} \right)   
        - \left( 2  r +1 \right)  \sqrt{\frac{1-r}{r} }
        \arctan \frac{\beta} {  \sqrt{\frac{1-r}{r} } }  \right\}  \right],
  \label{eq_for_mV}
\end{equation}
where we have defined a dimensionless coupling constant 
$\widetilde G_2 =  { G_2 N \Lambda^2}/{ 4 \pi^2}$ 
and $r$ is (square of) the ratio 
of the vector meson mass to the threshold mass $2M$:
\begin{equation}
   r \equiv \left( \frac{ m^{}_{\rm V}}{2M} \right)^2.
 \label{ratio}
\end{equation}
A physical bound-state appears only when the ratio $r$ is in the range 
$ 0 < r < 1$. Equation (\ref{eq_for_mV}) has a solution in this region 
when the strength of the coupling constant $\widetilde G_2$ 
is in the range $
    \widetilde G_2^{\rm{(min)}} < \widetilde G_2 <  \widetilde 
G_2^{\rm{(max)}}$ 
defined by
\begin{eqnarray}
    \widetilde G_2^{\rm{(min)}}  \equiv
   \frac{3}{2}  \left\{ \beta  + (1- \beta^2)
        \ln \left( \frac{1+\beta}{1-\beta} \right) \right\}^{-1},\quad
   \widetilde G_2^{\rm{(max)}}  \equiv 
\frac{3}{2} \cdot \frac{1}{\beta^3 }\, .
\end{eqnarray}
Two limiting cases $ \widetilde G_2 = \widetilde G_2^{\rm{(min)}} $ and 
$ \widetilde G_2 =\widetilde G_2^{\rm{(max)}} $ 
correspond to $r=1$ (loose binding limit), and $r=0$ (tight binding limit), 
respectively. When $M/ \Lambda \rightarrow 0\ (\beta\to 1)$, the physical 
bound-state region shrinks $\widetilde G_2^{\rm{(min)} } \rightarrow 
\widetilde G_2^{\rm{(max)} } $, while it becomes wider as $M/ \Lambda $ 
grows large. The existence of $\widetilde G_2^{\rm (min)}$ is consistent 
with the observation that there is no bound state in the NJL model
without the vector interaction. 
Similar behaviors have been found in Ref.~\cite{Dmitra}.

In Figure 1, a numerical solution to Eq.~(\ref{eq_for_mV}) 
is shown as a function 
of $\widetilde G_2$, where the constituent quark mass is taken to be 
$M/\Lambda=0.4\ (\beta=0.82)$ as an example. 
As we expect, the bound state appears for $\widetilde G_2$ larger than 
some critical value and the mass starts to decrease from the threshold value 
$2M$ as one increases the strength 
of the vector interaction. The value of critical coupling constants 
are exactly the same as the values predicted by analytic calculation. 
When $\beta=0.82$, they are $\widetilde G_2^{\rm{(min)}}=0.95, \ 
\widetilde G_2^{\rm{(max)}}=2.72$.

\begin{figure}
 \epsfxsize=10cm
 \centerline{\epsffile{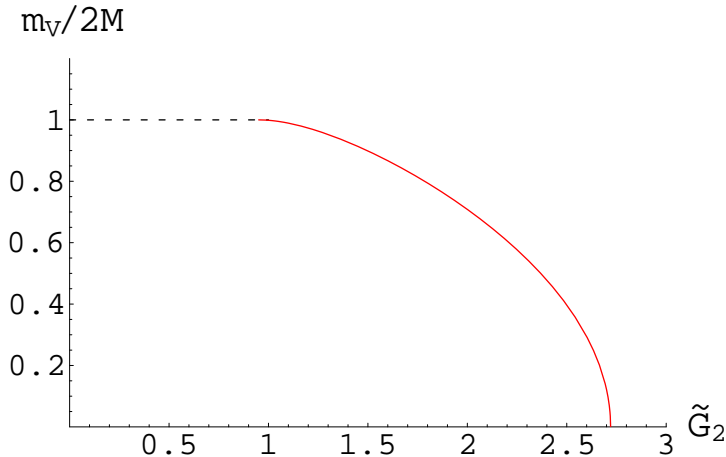}}
 \caption{$m_{\rm V}/2M$ as a function of $\widetilde G_2$. 
Constituent quark mass is $M/\Lambda=0.4$. A bound state appears only in 
the regime $\widetilde G_2^{\rm{(min)}}=0.95 \ <\widetilde G_2 < 
\widetilde G_2^{\rm{(max)}}=2.72$.}   
 \label{FIG1}
\end{figure}

It is also interesting to see $m^{}_{\rm V}$ as a function of $G_1$ and $G_2$.
The dependence on $G_1$ enters only through the constituent mass $M$ 
[see Eq.~(\ref{gap})]. 
In Figure 2, we show the vector meson mass $m^{}_{\rm V}$ as a function of 
$\widetilde G_1=G_1N\Lambda^2/4\pi^2$ and $\widetilde G_2$ in the broken 
phase $\widetilde G_1 > \widetilde G_1^{\rm (critical)}=1/2$ 
and in the chiral limit. 
As the coupling constant $\widetilde G_1$ becomes large, the constituent mass
(namely, the chiral condensate) becomes large. For fixed cutoff $\Lambda$,
this means to increase the value $M/\Lambda$ and thus enlarges the bound-state
region. This can be seen clearly in the figure.

\begin{figure}
 \centerline{\epsffile{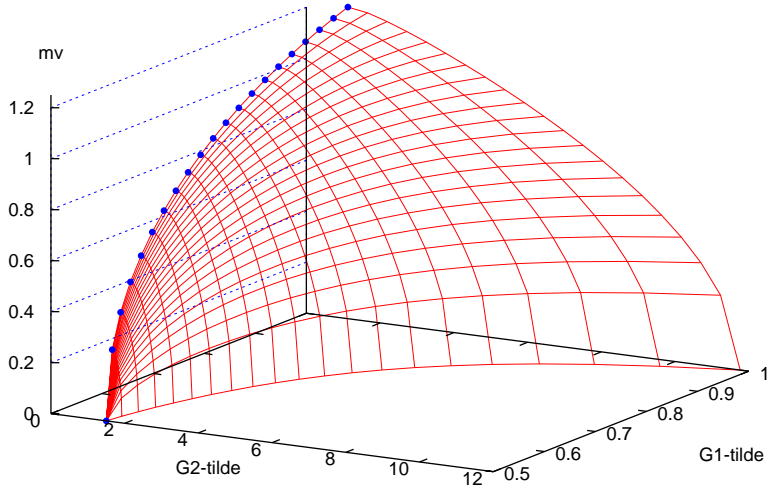}}
 \caption{Vector meson mass $m^{}_{\rm V}/\Lambda$ as a function of 
$\widetilde G_1$ and $\widetilde G_2$.}
 \label{FIG2}
\end{figure}

In Figure 3, we compare the LC wavefunction of 
the vector meson (\ref{EQ:SOL}) with that of the pseudo-scalar meson
$\phi_{\rm PS}(x,\kk)\propto (m_{\rm PS}^2-(\kk^2+M^2)/x(1-x))^{-1}$
which is the result of the previous analysis where the vector interaction 
was not included \cite{Itakura-Maedan}.
The transverse momentum $\kk$ is set to be zero for simplicity
and the wavefunctions are normalized at $x=1/2$ for comparison\footnote{{ 
If one includes the Lepage-Brodsky cutoff into the definition of 
the LC wavefunction,  support of $x$ for $\kk=0$ is given by 
$y_-<x<y_+$ with $y_\pm$ given by eq.~(\ref{y_beta}).}}.
We have chosen the masses to be $(m_{\rm PS}/2M)^2=0.01$ and 
$(m_{\rm V}/2M)^2=0.95$ as a typical case in the chirally broken phase 
with non-zero current quark mass $m^{}_0\neq 0$. 
In spite of the absence of the vector interaction in the previous analysis,
this comparison makes sense because the effects of the vector 
interaction on the mass of a pseudo-scalar meson is small. 
Indeed, after the Fierz 
transformation, the vector interaction will generate terms 
proportional to the original $G_1$ interaction, but of the 
higher order in $1/N$. Notice that the LC wavefunction of 
the pseudo-scalar meson has the same functional form of $x$ and $\kk$
as that of Eq.~(\ref{EQ:SOL}). Therefore, difference of the shape is 
due to different values of the mass. 
The constituent picture works better in the vector meson 
than in the pseudo-scalar meson.

\begin{figure}
 \epsfxsize=8cm
 \centerline{\epsffile{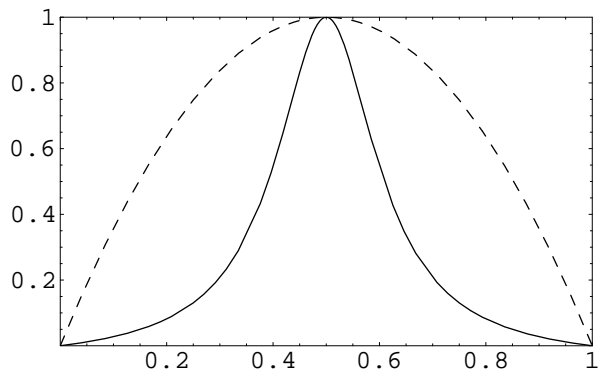}}
 \caption{{ LC wavefunctions $\phi(x,\kk=0)$ of pseudo-scalar (dashed) and 
vector (solid) mesons with typical masses 
$(m_{\rm PS}/2M)^2=0.01,\ (m_{\rm V}/2M)^2=0.95$.}
Normalized at $x=1/2$ for comparison.}
 \label{FIG3}
\end{figure}

Before concluding the paper, 
let us show one more evidence for the equivalence 
between the transverse and longitudinal equations. 
As we already mentioned, the superficial difference 
comes from the fact that the Lorentz covariance (in particular, the 
3 dimensional rotational invariance) is not manifest in the LF coordinates. 
If one works in a framework with obvious rotational invariance,
then there should be no difference between transverse and longitudinal 
components. 
Indeed, in the leading order of the $1/N$ expansion, it is possible to 
derive the same equations from the covariant Bethe-Salpeter (BS) equation:
\begin{equation}
 \hat{\phi}{}^{\rm BS}(q;P)
 = \frac{G_2 N}{(2\pi)^4 i} \int d^4 k\ {\rm tr}
 \left[\, \gamma^i \, \frac{1}{\Slash{k}-M+i \epsilon}\, \gamma_i\, 
 \hat{\phi}^{\rm BS}(k;P)\, \frac{1}{\Slash{k}-\Slash{P}-M-i\epsilon}\, 
 \right],
 \label{LCEQ:H150417:1}
\end{equation}
where $\hat{\phi}{}^{\rm BS}(q;P)$ is the amputated BS amplitude for the 
vector channel
($P$ and $q$ are the total momentum and relative momentum of a quark and an 
antiquark system, respectively). The LC wavefunction is obtained 
by the LF energy integral of the BS amplitude. 
Here, however, we have to be much careful because we must specify the 
cutoff scheme to make the equations well-defined.
When we impose the cutoff, it is of course (and again) 
desirable to maintain the 
symmetries such as the rotational invariance. For example, in 
Eq.~(\ref{LCEQ:H150417:1}), we can use the so called
3-momentum cutoff scheme which respects 3-dimensional rotational invariance.
Then, it turns out that the resulting two equations are equivalent 
to Eqs.~(\ref{EQ:TRANS}) and (\ref{EQ:LONG}) respectively, with the 
parity invariant cutoff (\ref{LB}) 
if one uses the following variable transformation 
$y  \leftrightarrow (E_k + {k_z})/2E_k,$ with 
$E_k=\sqrt{\Kmbf{k}^2+M^2}$ and $\Kmbf{k}$ being the three momentum here. 
Also, one needs to replace the 3-momentum cutoff $\Lambda_{\rm 3M}$ by the extended 
parity invariant cutoff $\Lambda$ using the relation shown before.
\\

To summarize, we have applied the LF quantization to the NJL model 
with the vector interaction, and 
obtained the eigenvalue equations for vector meson's LC wavefunctions.
Due to the addition of the vector interaction, the vector state 
becomes a bound state. At first glance, transverse and longitudinal 
components of the bound-state equations look different from each other, 
but eventually after imposing an appropriate cutoff scheme, one 
finds these two coincide with each other. Mass of the vector meson 
decreases as one increases the strength of the vector interaction.
This behavior is consistent with Refs.\cite{Dmitra,Kugo} which also treated 
the vector meson within the NJL model with the vector interaction. 

 Once we obtain the LC wavefunctions, we can compute various physical
quantities. One of such important quantities is the physical form factors.
For the pseudo-scalar case, this was done \cite{Heinzl} in the NJL model 
with two flavor quarks. Similar analysis can be done in the vector meson 
case and will be reported in the future publications \cite{full_paper}.

Our non-perturbative approach plays a complementary role to the perturbative 
calculation of the asymptotic form \cite{Braun-Koike-Tanaka}, because 
our LC wavefunction is expected to describe that of low energy scale 
(Of course we have to include flavor degrees of freedom, which 
is straightforward. See for example, \cite{Bentz}). It is 
interesting to find a way to interpolate these two different approaches. 
One of the possible ways for this problem is to include the effects of 
gluon propagation between a quark and an antiquark. 
This is easily incorporated 
by using the nonlocal current current interaction $j^\mu D_{\mu\nu}j^\nu$ 
instead of the point interaction in the NJL model. Such kind of interpolation 
between low and high momentum regimes is used in various situations.
It is also interesting to apply our approach to the heavy quark system
such as $J/\psi$ or $\Upsilon$.
The use of the nonlocal interaction is also convenient from the technical 
point of view. In the presence of the gauge field, one needs to consider 
the longitudinal zero modes of the gauge field. In spite of the favorable 
aspect that the nontrivial vacuum structure such as the theta vacua may 
be attributed to the dynamics of gauge zero modes, inclusion of the 
gauge zero modes makes the canonical structure terribly complicated 
\cite{gaugeLF}. Thus, for the 
problem of quarkoniums where one can ignore the vacuum physics, it is 
easier to replace the gauge field effects by the non-local current-current 
interaction. \\


\end{document}